\documentstyle[preprint,aps,epsfig]{revtex}
\begin{document} 
\preprint{Phys. Rev. C, Rapid Communication, (2001) in press.}  
\title{Isospin Dependence of Mechanical and Chemical Instabilities in 
Neutron-Rich Matter} 
\bigskip 
\author{\bf Bao-An Li, Andrew T. Sustich, Matt Tilley and Bin Zhang} 
\address{Department of Chemistry and Physics\\
P.O. Box 419, Arkansas State University\\
State University, AR 72467-0419, USA}
\maketitle

\maketitle 
\begin{quote} 
Within nuclear thermodynamics and an isospin-dependent transport model we 
investigate respective roles of the nuclear mean field and the 2-body stochastic 
scattering on the evolution of density and isospin fluctuations in either 
mechanically or chemically unstable regions of neutron-rich matter. It is found 
that the mean field dominates overwhelmingly the fast growth of both fluctuations, 
while the 2-body scattering influences significantly the later growth of the 
isospin fluctuation only. Moreover, both fluctuations grow in mechanically unstable 
systems, while only the density fluctuation grows significantly in chemically unstable 
ones. Furthermore, the magnitude of both fluctuations decreases with the increasing 
isospin asymmetry because of the larger reduction of the attractive isoscalar mean 
field by the stronger repuslive neutron symmetry potential in the more neutron-rich matter.
Finally, several experimental measurements are proposed to test these findings.
\end{quote}  
\newpage 
The rapid advance in experiments with rare isotopes has opened up several 
new frontiers in nuclear science\cite{rib}. The planned Rare Isotope 
Accelerator ({\rm RIA})will further enhance the exploration of these forefronts 
dramatically\cite{ria}. Besides detailed information on the structure of exotic 
nuclei in many unexplored regions of the periodic chart, it is now possible to 
study novel properties of isospin-asymmetric nuclear matter with extreme 
neutron to proton ratios. Prospects for discovering new physics in warm 
neutron-rich matter of various densities that can be created transiently in 
nuclear reactions induced by exotic neutron-rich nuclei have generated much 
interest in the nuclear science 
community\cite{li01,li97,li98,dasgupta,tsang01,gary,sherry}. In particular, 
the density dependence of the symmetry term in the equation of state ({\rm EOS})
has recently received much attention since it is among the most important but very poorly 
known properties of neutron-rich matter\cite{wir88,brown00,hor00,dito01,lom01,bom01}. 
This term is very important to the mechanisms of Type II supernova 
explosions and neutron-star mergers. It also determines the proton fraction 
and electron chemical potential in neutron stars at $\beta$ equilibrium. These 
quantities consequently influence the cooling rate of protoneutron stars and 
the possibility of kaon condensation in dense stellar 
matter\cite{lat91,bom94,sum94,lee96,pra97}. 

Based on the {\rm EOS} obtained from various microscopic many-body theories 
and phenomenological models\cite{lat78,bar80,muller,liko97,baran98,cat01}, 
it has long been predicted that isospin-asymmetric nuclear matter under certain 
conditions can be mechanically or chemically unstable, i.e.,  
\begin{equation}  
\left(\frac{\partial P}{\partial \rho}\right)_{T,\delta}\leq 0~~~({\rm mechanical}),
\end{equation}
or
\begin{equation}
\left(\frac{\partial \mu_n}{\partial \delta}\right)_{P,T}\leq 0~~~({\rm chemical}), 
\end{equation} 
where $P$, $\mu_n$ and $\delta=(\rho_n-\rho_p)/(\rho_n+\rho_p)$ are the pressure, 
neutron chemical potential and isospin asymmetry, respectively.  
In the mechanically or chemically unstable regions, small fluctuations in 
density and/or isopin-asymmetry are generally expected to grow. However, information 
about the respective growth rates of these fluctuations, how and why they may depend on 
the isospin-asymmtry of nuclear matter, seeds of fluctuations and their isospin-dependence,
how and which aspects of nuclear dynamics cause the growth of these fluctuations 
is rather rare. This information is vitally important for understanding 
the structure and stability of both neuton stars and radioactive nuclei as well 
as mechanisms of nuclear multifragmentation in reactions induced by neutron-rich nuclei. 
In this Rapid Communication, we obtain such information within nuclear thermodynamics and an isospin-dependent 
transport model. We also propose several experimental measurements to test our findings.

The boundaries of the mechanical and chemical instabilities in the configuration 
space of density $\rho$, isospin asymmetry $\delta$ and temperature $T$ can
be established readily by using a thermal model for isospin asymmetric nuclear 
matter\cite{liko97}. We use here a Skyrme-type phenomenological {\rm EOS} for 
isospin-asymmetric nuclear matter in the parabolic approximation\cite{bom91,hub93}.
The energy per nucleon at $T=0$ can be written as 
\begin{equation}\label{aeos}
e(\rho,\delta)= \frac{a}{2}+\frac{b}{1+\sigma}u^{\sigma}+\frac{3}{5}e_F^0u^{2/3}
+S_0(\rho_0)\cdot u^{\gamma}\cdot\delta^2.
\end{equation}
In the above $u\equiv \rho/\rho_0$ is the reduced density, $e_F^0$ is the Fermi energy
in symmetric nuclear matter at normal density,  and  
$a=-123.6$ MeV, $b=70.4$ MeV and $\sigma=2$ corresponding to a stiff 
nuclear {\rm EOS} of isospin-symmetric nuclear matter. The last term is the 
symmetry energy whose density dependence is currently rather 
uncertain\cite{wir88,brown00,hor00,dito01,lom01,bom01}. We adopt
here a parameterization used by Heiselberg and Hjorth-Jensen in their recent 
studies of neutron stars\cite{hei00} with $S_0(\rho_0)=30$ MeV and $\gamma$ 
as a free parameter. We note that a value of about $\gamma=0.6$ was obtained
by fitting to the result of variational many-body calculations\cite{hei00,akm97}. 
The nucleon chemical potential in asymmetric 
nuclear matter $\mu_q$,(q=n for neutrons, q=p for protons) at temperature $T$ 
is thus\cite{jaqaman1}
\begin{equation}\label{muq} 
\mu_q=au+bu^{\sigma}+v_{\rm asy}^{q}+T\left[{\rm ln}(\frac{\lambda_T^3\rho_q}{2}) 
+\sum_{n=1}^{\infty}\frac{n+1}{n}b_n(\frac{\lambda_T^{3}\rho_q}{2})^n\right], 
\end{equation} 
where $\lambda_T=\left[2\pi\hbar^{2}/(m_q T)\right]^{1/2}$ is the thermal 
wavelength of a nucleon. The coefficients $b_n$ are obtained from mathematical 
inversion of the Fermi distribution function\cite{jaqaman1}. 
The single particle symmetry potential $v_{\rm asy}^q$ is
\begin{equation}\label{vasy}
v_{\rm asy}^{q}=\pm 2\left[S_0 \cdot u^{\gamma}-12.7u^{2/3}\right]\delta
+\left[S_0(\gamma-1)u^{\gamma}+4.2u^{2/3}\right]\delta^2,
\end{equation}
where ``+" and ``-" are for neutrons and protons, respectively.
The corresponding pressure is
\begin{eqnarray} 
P&=&\frac{1}{2}a\rho_0u^{2}+b\frac{\sigma\rho_0}{\sigma+1}u^{\sigma+1} 
+(S_0\gamma\rho_0u^{\gamma+1}-8.5\rho_0u^{\frac{5}{3}})\delta^{2}\nonumber\\
&+&T\rho\{1+\frac{1}{2}\sum_{n=1}^{\infty} 
b_n(\frac{\lambda_T^{3}\rho}{4})^n\left[(1+\delta)^{n+1} 
+(1-\delta)^{n+1}\right]\}. 
\end{eqnarray} 
We then use the Gibbs-Duhem relation 
\begin{equation}\label{gibbs} 
\frac{\partial P}{\partial \rho}=\frac{\rho}{2}\left[(1+\delta) 
\frac{\partial \mu_n}{\partial \rho}+(1-\delta)
\frac{\partial \mu_p}{\partial \rho}\right] 
\end{equation} 
to find boundaries of mechanically unstable regions 
(ITS: isothermal spinodal) in the
$\rho-\delta$ plane for given temperatures. 
Since the chemical instability condition has to be evaluated at constant pressures, 
the following Maxwellian relation 
\begin{equation}\label{trans} 
\left(\frac{\partial \mu_n}{\partial \delta}\right)_{T,P}
=\left(\frac{\partial \mu_n}{\partial \delta}\right)_{T,\rho}
-\left(\frac{\partial \mu_n}{\partial \rho}\right)_{T,\delta} 
\cdot\left(\frac{\partial P}{\partial \rho}\right)^{-1}_{T,\delta} 
\cdot\left(\frac{\partial P}{\partial \delta}\right)_{T,\rho} 
\end{equation} 
is used to find boundaries of the chemically unstable 
regions (DS: diffusive spinodal).

Shown in Fig.\ 1 are the boundaries of the mechanical (thick lines) and 
chemical (thin lines) instabilities in the $\rho-\delta$ plane with $\gamma=0.5$ 
at $T=5$ (lower window), 10 (middle window) and 15 MeV (upper window), 
respectively. It is seen that the diffusive spinodal extends further out into 
the plane and envelopes the region of mechanical instability; the two regions of 
instability do no overlap. As the temperature increases, both instabilities
become less prominent over a more narrow range of densities at smaller isospin 
asymmetries. These features are independent of the parameters of the {\rm EOS} 
and are in good agreement with those based on more microscopic many-body 
theories\cite{lat78,bar80,muller}. In particular, we found that the variation of 
$\gamma$ parameter has very little effect on the instability boundaries. 
Therefore, in the following a constant of $\gamma=0.5$ is used. The results 
shown in Fig.\ 1 not only provide the motivations but also serve as a guidance
for our following numerical simulations.

The evolution of isospin-asymmetric nuclear matter in the mechanically or 
chemically unstable regions is studied by using the isospin-dependent transport 
model\cite{li97,li98}. The model uses consistently the isospin-dependent 
{\rm EOS} and the corresponding potentials as in the thermal model outlined above. 
Moreover, scatterings among neutrons and protons are fully isospin-dependent in 
terms of their total and differential cross sections as well as the Pauli 
blockings. Nucleons are initialized in momentum space using a Boltzmann distribution 
function in a cubic box of length $L_{box}$ with periodic boundary conditions. We 
have checked that the use of Fermi-Dirac distribution does not alter our results at 
all as we are mainly interested in the evolution of reduced 
fluctuations with respect to the initial state which is unstable against numerical 
fluctuations in the mechanical or chemical instability region. Thus, the initialization 
with the Boltzmann distribution is sufficient and much more efficient numerically. 
The box is further divided into cells of 1 $fm^3$ volume
in which the average density $\rho_{cell}$ and isopsin asymmetry $\delta_{cell}$
are evaluated. We used $10^{4}$ test particles per nucleon in evaluating the
$\rho_{cell}$ and $\delta_{cell}$. Shown in Fig. 2 is an illustration of the 
evolution of a system initialized in the mechanically unstable region with 
$T_i=5$ MeV, $\delta_i=0.6$ and $\rho_i=0.05$ $fm^{-3}$. 
The most interesting feature shown here is the gradually increasing isospin 
fractionation. This is indicated by the spreading of the initial system into 
regions with $\rho\leq \rho_i$ and $\delta\geq\delta_i$ and where 
$\rho\geq \rho_i$ but $\delta\leq\delta_i$. The variations of $\rho$ 
and $\delta$ with respect to their initial values can be characterized 
quantitatively by using, respectively, $\sigma_d(t)=(\bar{\rho^2}-\rho_i^2)^{1/2}$ 
and $\sigma_{\delta}(t)=(\bar{\delta^2}-\delta_i^2)^{1/2}$, where the average is 
over all cells. Furthermore, the degree of isospin fractionation can be quantified
by using the ratio $(N/Z)_{gas}/(N/Z)_{liquid}$, where $(N/Z)_{gas}$ and $(N/Z)_{liquid}$
is the isospin asymmetry of the low $(\rho/\rho_0\leq 1/8)$ and 
high $(\rho/\rho_0> 1/8)$ density region, respectively. 

Shown in Fig.\ 3 are the reduced variation with respect to the initial state 
in isospin asymmetry $\sigma_{\delta}(t)-\sigma_{\delta}(0)$, density $\sigma_d(t)/\rho_i$ 
and the degree of isospin fractionation $(N/Z)_{gas}/(N/Z)_{liquid}$
as a function of time for a system initialized 
at $T_i=5$ MeV, $\rho_i=$0.05 $fm^{-3}$ and $\delta_i=$0.2, 0.6 and 0.9,
respectively. As a reference and check of our approach, results (dash-dot lines) are also shown for a 
system initialized with $T_i=15$ MeV and $\delta_i=0.9$ where it is both 
mechanically and chemically stable. For this system, it is seen that both 
the isospin and density fluctuations stay almost constant and there is no 
isospin fractionation at all. While for the system initialized in  
the mechanically (chemically) unstable region with $\delta_i=$0.2 
and 0.6 ($\delta_i=$0.9), it is seen that both the isospin and density fluctuations 
grow {\it faster} with the {\it decreasing} isospin asymmetry $\delta_i$, 
and the opposite trend is observed for the strength of isospin fractionation.
Moreover, both the density and isospin fluctuations grow in the mechanically unstable 
systems, while only the density fluctuation grows significantly in the chemically 
unstable ones. It is also interesting to note that the isospin fractionation happens later
in the more neutron-rich matter. In the latter early fluctuations are smaller and 
thus take longer time to grow before they can trigger the isospin fractionation.

We now proceed to explore the seeds of fluctuations and their isospin dependence, 
and to investigate how and which aspects of nuclear dynamics are important in governing the 
growth of fluctuations. We shall also study the question why the more neutron-rich matter is 
more stable against both the density and isospin fluctuations. The evolution dynamics 
of asymmetric nuclear matter is governed by the isospin-dependent nuclear mean filed 
and stochastic nuclear scatterings. In our approach, first-order effects of 2-body 
stochastic scatterings are included through the collision 
integral of the BUU equation. Therefore, there are two main seeds of fluctuations in our approach, 
i.e., the early numerical fluctuations 
from the random sampling of the initial state and the later 2-body stochastic nuclear scatterings. 
Both of them may lead to the growth of fluctuations by propagating  
through the nuclear mean field in the early and later stages of the evolution, respectively. 
Shown in the upper window of Fig.\ 4 are the average number of collisions per nucleon as a function of 
time for $\delta=0.2$ and 0.9, respectively. Before about 40 fm/c, there is essentially no collision 
because of the strong Pauli blocking when the phase space nonuniformity due to the initial fluctuation 
is still small. Later, nuclear scatterings become important, moreover, they are more frequent with 
the decreasing isospin asymmetry. The less frequent 2-body scatterings observed with the higher 
$\delta$ is because of the isospin-dependence of both the nucleon-nucleon cross sections and the
Pauli blocking rates. It is well known that the cross section for neutron-neutron scatterings is 
only about $1/3$ of that for neutron-proton collisions at beam energies below about 1 GeV\cite{nndata}.  
One also expects from Fermi statistics that neutron-neutron scatterings are more strongly Pauli blocked in the more 
neutron-rich matter. To investigate the respective roles of the nuclear mean field and the 2-body scatterings, 
model studies by turnning off the nucleon-nucleon collisions have been performed. 
As shown in the middle window of Fig.\ 4,  the collisional seeds of fluctuations 
lead to the significant growth of the isospin fluctuation in the later stage of the evolution.
However, they have very little effect on the growth of density fluctuations as shown in the lower window. 
A comparison of the results obtained with and without the collision integral 
indicates that the fluctuations are overwhelmingly dominated by the nuclear mean field.
The observed isospin-dependence of the fluctuations can be understood from the interplay 
between the {\it attractive isoscalar} mean field and the {\it repulsive symmetry potential for neutrons}. 
The symmetry potential, as shown in Eq.\ \ref{vasy}, is repulsive for neutrons and attractive for protons 
and their magnitudes increase with the increasing isospin asymmetry. Thus, the resultant attractive mean 
field is weaker in the more neutron-rich matter. Because of the small number of scatterings, particularly 
in the early stage of the evolution, the growth of fluctuations is thus mainly determined by the strength and 
sign of the resultant nuclear mean field according to the linear response theory for asymmetry nuclear 
matter\cite{baran98,baran01,ditoro}. Based on the latter, the magnitude of both fluctuations can grow larger 
with the increasing strength of the attractive resultant mean field, and thus also with the decreasing 
isospin asymmetry $\delta$.

What are the important physical implications of our findings? We expect several 
effects that can be tested experimentally.
It is well known that more neutron-rich systems are less bound and have smaller
saturation densities. However, fluctuations and their growth are also important 
for determining the final state of a neutron-rich system, such as in the 
projectile fragmentation in producing exotic beams\cite{fri00}. Our results above 
indicate that fluctuations actually have a compensating role to the lower 
binding energy in stablizing neutron-rich system where fluctuations  
are smaller and do not grow as fast as in symmetric ones. 
Moreover, our results on the isospin fractionation accompanying the evolution 
of fluctuations indicate that the configuration of a more 
dense, isopin-symmetric region surrounded by a more isospin-asymmetric gas 
as in halo nuclei is a natural result of the isospin-dependent nuclear dynamics. 
Furthermore, we expect that the multifragmentation and isospin fractionation 
in nuclear reactions induced by neutron-rich nuclei to happen on longer time 
scales compared to symmetric reactions of the same masses. These expectations 
can be tested by measuring products of multifragmentation, in particular, 
the neutron-neutron, proton-proton as well as fragment-fragment correlation 
functions, in comparative studies of isospin symmetric and asymmetric nuclear 
reactions\cite{fri00,robert,bauer,moretto}. 

In summary, we have investigated the isospin dependence of mechanical and 
chemical instabilities in neutron-rich matter within nuclear thermodynamics and 
the isospin-dependent transport model. We found that the mean field dominates 
overwhelmingly the fast growth of both density and isospin fluctuations, 
while the 2-body scatterings influence significantly the later, slower 
growth of the isospin fluctuation only. Moreover, both of the fluctuations 
grow in the mechanically unstable systems, while only the density fluctuation 
grows significantly in the chemically unstable ones. Furthermore, the magnitude 
of both fluctuations decreases with the increasing isospin asymmetry 
because of the larger reduction of the attractive isoscalar nuclear 
mean field by the stronger repuslive neutron symmetry potential in the more 
neutron-rich matter. Finally, several experimental measurements are 
proposed to test these findings.

We would like to thank Amanda Evans, Betty Tsang and S.J. Yennello for useful discussions. 
This work was supported in part by the National Science Foundation Grant No. 0088934 and 
Arkansas Science and Technology Authority Grant No. 00-B-14.

\newpage 
\begin{figure}[htp] 
\centering \epsfig{file=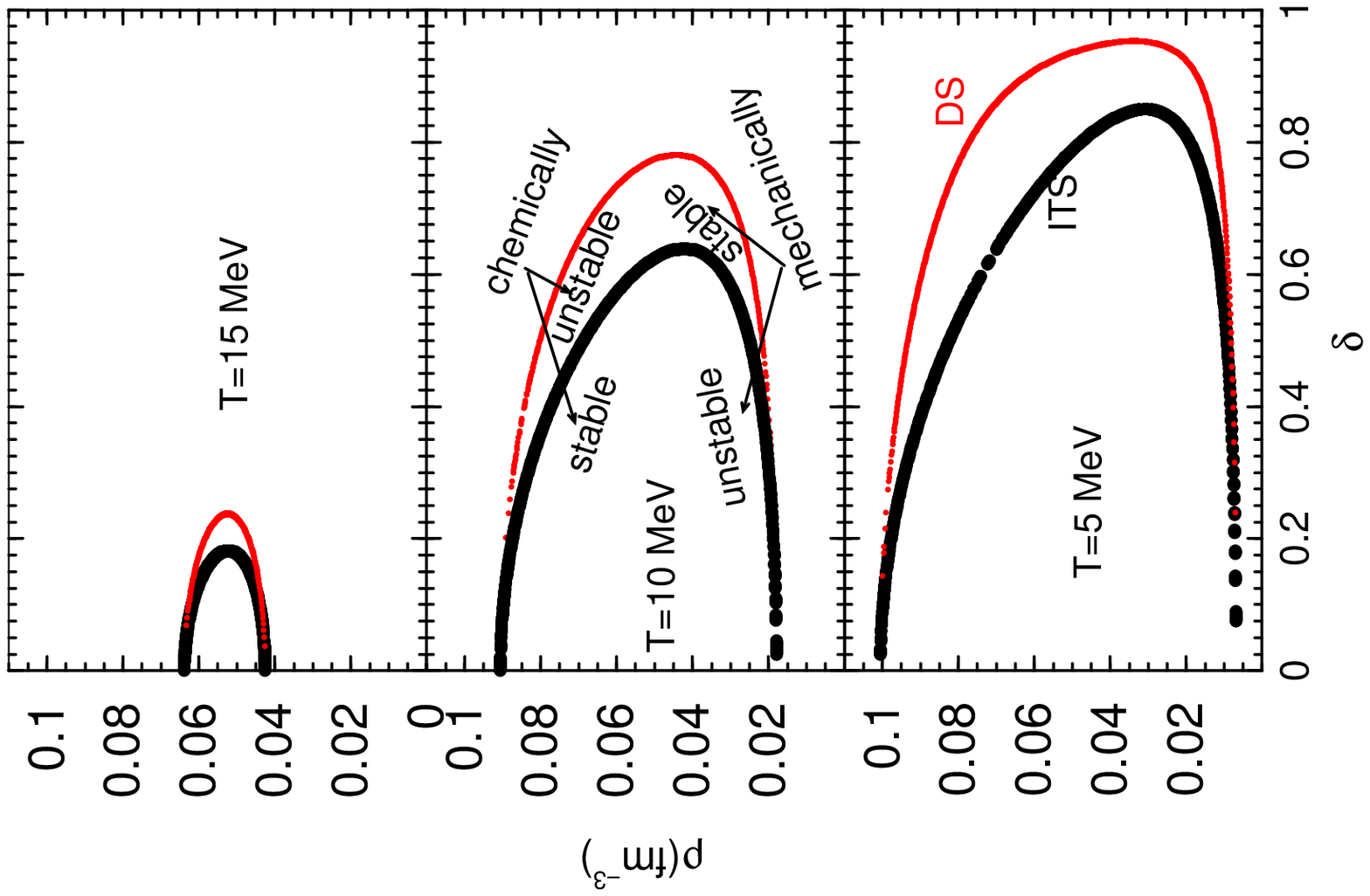,width=15cm,height=12cm,angle=-90} 
\caption{Boundaries of the mechanical (thick lines) and chemical (thin lines)
in the density-isospin asymmetry plane at a temperature of 5, 10 and 15 MeV, 
respectively. } 
\end{figure} 

\begin{figure}[htp] 
\centering \epsfig{file=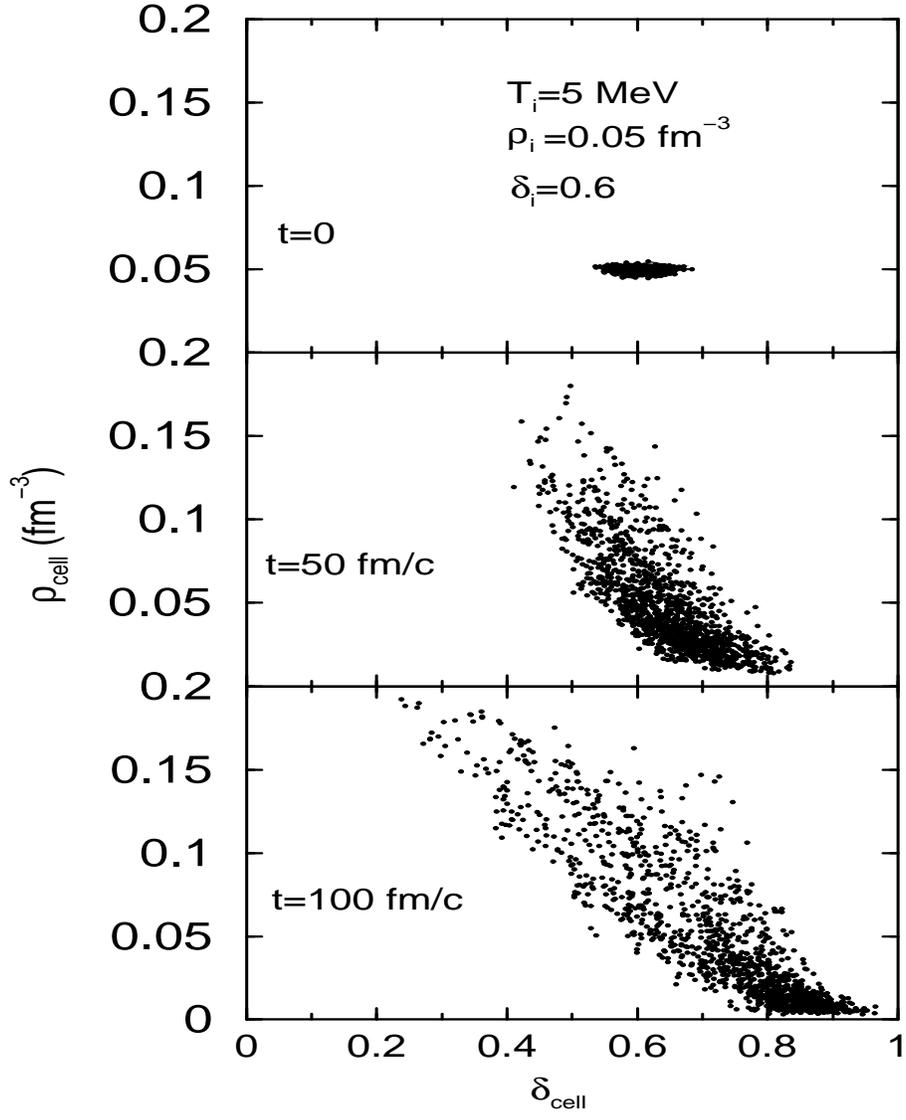,width=15cm,height=12cm,angle=-90} 
\caption{An illustration of the evolution of asymmetric nuclear matter
in the density-isospin asymmetry plane. Each point in the scatter plots 
represent one cell of 1 $fm^3$ volume in a cubic box of side length 
$L_{box}=10$ fm.}
\end{figure}   

\begin{figure}[htp] 
\centering \epsfig{file=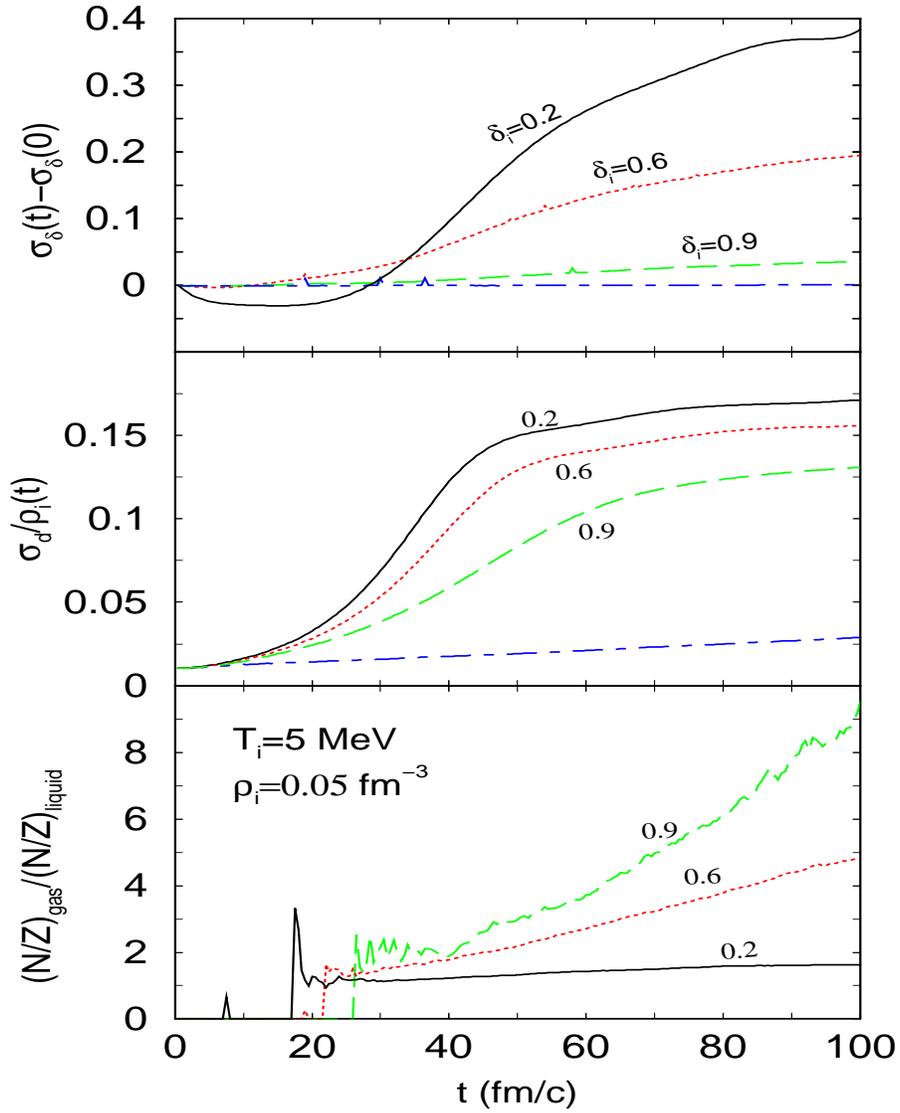,width=15cm,height=12cm,angle=-90} 
\caption{Evolution of the reduced isospin fluctuation (upper window),
reduced density fluctuation (middle window) and strength of 
isospin fractionation (lower window)in a cubic box of side 
length $L_{box}=30$ fm and density 0.05 $fm^{-3}$. The dash-dot lines 
are calculated with $T_i=15$ MeV and $\delta_i=0.9$; while the solid, dot and 
dashed lines are calculated with $T_i=5$ MeV and $\delta_i=$0.2, 0.6 and 0.9, 
respectively.}   
\end{figure}  

\begin{figure}[htp] 
\centering \epsfig{file=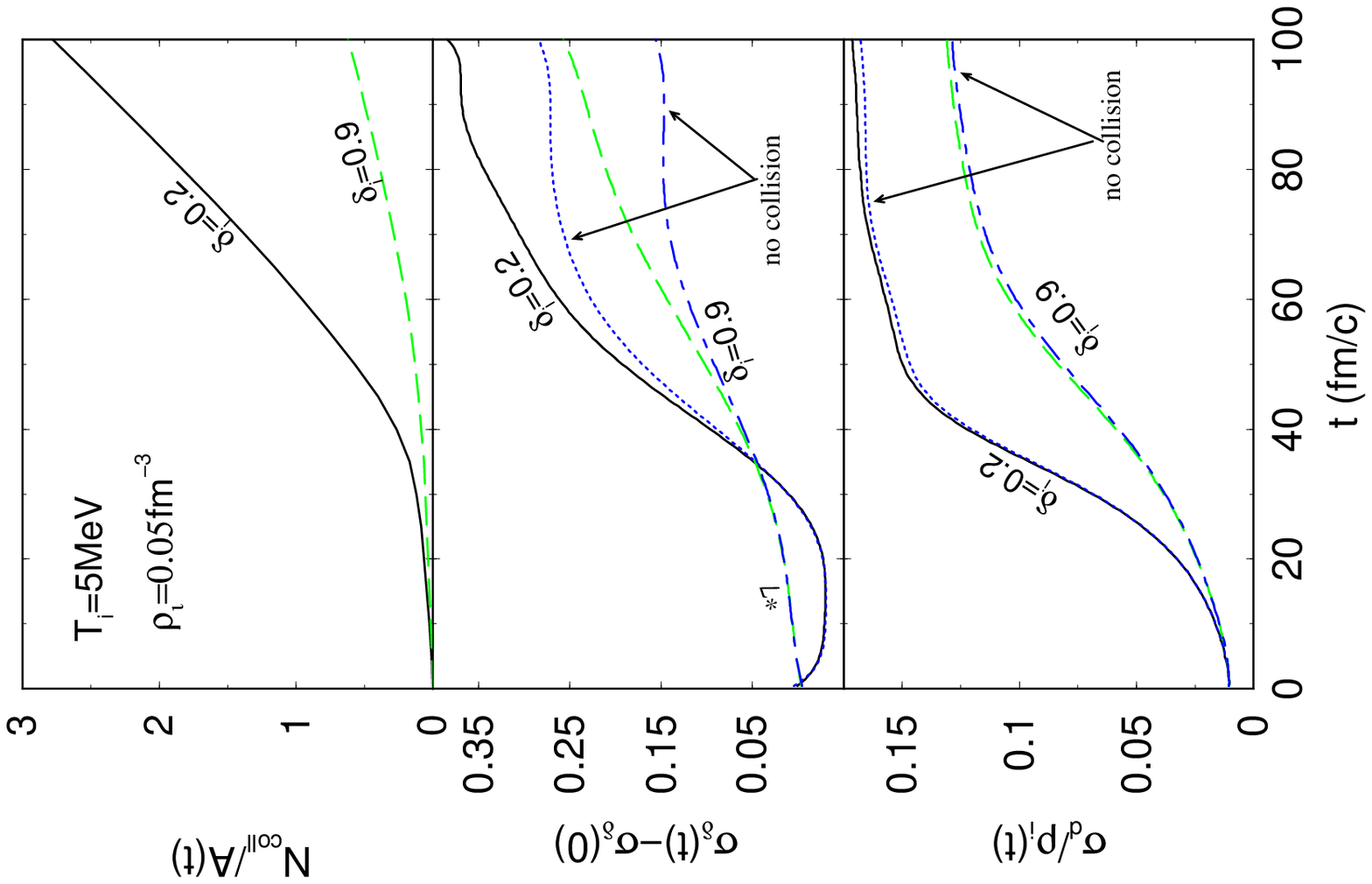,width=15cm,height=12cm,angle=-90} 
\caption{The average number of successful nucleon-nucleon collisions per nucleon (upper window),
the isospin (middle window) and density (lower window) fluctuations 
as a function of time. The initial conditions are the same as in Fig. 3. The dot and dash-dot lines
are results of calculations without the 2-body scatterings. For comparisions on the same scale, a factor of 7
is multiplied to the isospin fluctuations with $\delta=0.9$ in the middle window. }
\end{figure} 


\newpage \begin{thebibliography}{99} 

\bibitem{rib}A. Mueller and B. Sherril, Annu. Rev. Nucl. Part. Sci.
{\bf 43}, 529 (1993); P.G. Hansen, A.S. Jensen and B. Jonson, Ann. Rev. Nucl.
Part. Sci. {\bf 45}, 591 (1995); I. Tanihata, Prog. of Part. and Nucl. Phys., 
{\bf 35} (1995) 505; W. Nazarewicz, B. Sherril, I. Tanihata and 
P. Van Duppen, Nucl. Phys. News {\bf 6}, 17 (1996).

\bibitem{ria}RIA Physics White Paper, 2000, Eds. R. Casten, W. Nazarewicz et al..

\bibitem{li01}Isospin Physics in Heavy-Ion Collisions at Intermediate Energies,
Eds. Bao-An Li and W. Udo Schr\"oder, ISBN 1-56072-888-4, 
Nova Science Publishers, Inc (2001, New York).

\bibitem{li97}B.A. Li et al., Phys. Rev. Lett. {\bf 76}, 4492 (1996); 
{\it ibid}, {\bf 78}, 1644 (1997); B.A. Li, Phys. Rev. Lett. {\bf 85}, 4221 (2000); B.A. Li, 
Nucl. Phys. {\bf A681}, 434c (2001).

\bibitem{li98}B.A. Li, C.M. Ko and W. Bauer, 
topical review, Int. Jou. Mod. Phys. E{\bf 7}, 147 (1998).

\bibitem{dasgupta}J. Pan and S. Das Gupta, Phys. Rev. C{\bf 57}, 1839 (1998)

\bibitem{tsang01}H.S. Xu et al., Phys. Rev. Lett. {\bf 85}, 716 (2000); 
M.B. Tsang et al, nucl-ex/0103010, Phys. Rev. Lett. (2001) in press. 

\bibitem{gary}G.D. Westfall, Nucl., Phys. {\bf A681}, 343c (2001).

\bibitem{sherry}S.J. Yennello et al., Nucl., Phys. {\bf A681}, 317c (2001).

\bibitem{wir88}R.B. Wiringa, V. Fiks and A. Fabrocini, 
Phys. Rev. C{\bf 38}, 1010 (1988).

\bibitem{brown00}B.A. Brown, Phys. Rev. Lett. {\bf 85}, 5296 (2000).

\bibitem{hor00}C.J. Horowitz et al., Phys.Rev. C{\bf 63}, 025501 (2001).

\bibitem{dito01}M. Colonna et al., Phys. Lett. {\bf B428}, 1 (1998). 

\bibitem{lom01}U. Lombardo and W. Zuo in ref. \cite{li01}.

\bibitem{bom01}I. Bombaci in ref. \cite{li01}.

\bibitem{lat91}J.M. Lattimer, C.J. Pethick, M. Prakash and P. Haensel, 
Phys. Rev. Lett. {\bf 66}, 2701 (1991).

\bibitem{bom94}I. Bombaci, T.T.S. Kuo and U. Lombardo, 
Phys. Rep. {\bf 242}, 165 (1994).

\bibitem{sum94}K. Sumiyoshi and H. Toki, Astro. Phys. Journal, {\bf 422}, 
700 (1994); K. Sumiyoshi, H. Suzuki and H. Toki,
Astronomy and Astrophysics, {\bf 303}, 475 (1995).

\bibitem{lee96}C.-H. Lee, Phys. Rep. {\bf 275}, 255 (1996).

\bibitem{pra97}M. Prakash et al., Phys. Rep. {\bf 280}, 1 (1997); 
J. Lattimer and M. Prakash, Phys. Rep. {\bf 333}, 121 (2000).

\bibitem{lat78}J.M. Lattimer and D.G. Ravenhall, 
Astr. Jour., {\bf 223}, 314 (1978). 

\bibitem{bar80}M. Barranco and J. R. Buchler, Phys. Rev. C{\bf 22}, 1729 (1980).

\bibitem{muller}H. M\"uller and B.D. Serot, Phys. Rev. C {\bf 52}, 2072 (1995).

\bibitem{liko97}B.A. Li and C.M. Ko, Nuc. Phys. A {\bf 618}, (1997) 498.   

\bibitem{baran98}V. Baran et al. , 
Nucl. Phys. {\bf A632}, 287 (1998).

\bibitem{cat01} D. Catalano, G. Giansiracusa and U. Lombardo, 
Nucl. Phys. {\bf A681}, 390c (2001).

\bibitem{bom91}I. Bombaci and U. Lombardo, 
Phys. Rev. C{\bf 44}, 1892 (1991).

\bibitem{hub93}H. Huber, F. Weber and M.K. Weigel, 
Phys. Lett. {\bf B317}, 485 (1993); Phys. Rev. C{\bf 50}, R1287 (1994).

\bibitem{hei00}H. Heiselberg and M. Hjorth-Jensen, 
Phys. Rep. {\bf 328}, 237 (2000).

\bibitem{akm97}A. Akmal and V.R. Pandharipande, Phys. Rev. C{\bf 56}, 2261 (1997);
A. Akmal, V.R. Pandharipande and D.C. Ravenhall, Phys. Rev. C{\bf 58}, 1804 (1988).

\bibitem{jaqaman1}H.R. Jaqaman, Phys. Rev. C{\bf 39}, 169 (1988); 
{\it ibid} C{\bf 40}, 1677 (1989).  

\bibitem{nndata}G. Alkahzov et al., Nucl. Phys. {\bf A280}, 365 (1977).

\bibitem{baran01} V. Baran et al., 
Phys. Rev. Lett. {\bf 86}, 4492 (2001).

\bibitem{ditoro}V. Baran et al, private communications.

\bibitem{fri00}W.A. Friedman, M.B. Tsang, D. Bazin and W.G. Lynch,
preprint MSUCL-1167, July 2000.

\bibitem{robert}R. Ghetti et al., Nucl. Phys. {\bf A674}, 277 (2000);
Phys. Rev. C{\bf 62}, 037603 (2000).

\bibitem{bauer}W. Bauer, C.K. Gelbke and S. Pratt, 
Ann. Rev. Nucl. Part. Sci. {\bf 42}, 77 (1992).

\bibitem{moretto}L.G. Moretto and G.J. Wozniak, 
Ann. Rev. Nucl. Part. Sci. {\bf 43} (1993) 123.  

\end{thebibliography}
\end{document}